\documentclass[aps,floatfix,preprint,nofootinbib]{revtex4}
\usepackage{graphics}
\usepackage{graphicx}
\usepackage{amssymb}		
\usepackage{amsmath}
\usepackage{verbatim}
\usepackage{pstricks}
\usepackage{slashed}

\newcommand{\beq}{\begin{equation}}
\newcommand{\eeq}{\end{equation}}
\newcommand{\bea}{\begin{eqnarray}}
\newcommand{\eea}{\end{eqnarray}}

\newcommand{\vev}[1]{\langle #1 \rangle}
\newcommand{\op}{\mathcal{O}}

\begin{document}

\setlength{\baselineskip}{0.22in}

\begin{flushright}MCTP-10-18 \\
\end{flushright}
\vspace{0.2cm}

\title{Asymmetric Dark Matter from a GeV Hidden Sector}

\author{Timothy Cohen, Daniel J. Phalen, Aaron Pierce, and Kathryn M. Zurek}
\vspace{0.2cm}
\affiliation{Michigan Center for Theoretical Physics (MCTP) \\
Department of Physics, University of Michigan, Ann Arbor, MI
48109}

\date{\today}

\begin{abstract}
Asymmetric Dark Matter (ADM) models relate the dark matter density to the baryon asymmetry, so that a natural mass scale for ADM is around a few GeV.  In existing models of ADM, this mass scale is unexplained; here we generate this GeV scale for dark matter (DM) from the weak scale via gauge kinetic mixing with a new Abelian dark force.  In addition, this dark sector provides an efficient mechanism for suppressing the symmetric abundance of DM through annihilations to the dark photon. We augment this sector with a higher dimensional operator responsible for communicating the baryon asymmetry to the dark sector.  Our framework also provides DM candidate for gauge mediation models.  It results in a direct detection cross section of interest for current experiments: $\sigma_p \lesssim 10^{-42} \mbox{ cm}^2$ for DM masses in the range $1-15 \mbox{ GeV}$. 
\end{abstract}

\maketitle

\section{Introduction}

Despite its successes, the Standard Model of particle physics (SM) fails to account for either the origin of the baryon--anti-baryon asymmetry or the identity of the Dark Matter (DM).   In the standard thermal freeze-out paradigm for DM, the DM and baryon densities are determined by unrelated dynamical processes.  By contrast, in the framework of Asymmetric Dark Matter (ADM)  
\cite{Kaplan:1991ah,Nussinov:1985xr,Barr:1991qn,Barr:1990ca,Gudnason:2006ug,Kuzmin:1996he,Dodelson:1991iv,Fujii:2002aj,Kitano:2004sv,Farrar:2005zd,Kitano:2008tk,Kaplan:2009ag,An:2009vq}, the number density of the dark matter is set by the baryon asymmetry.  Because the observed DM energy density is comparable to the baryon density today, the DM mass in these 
models is usually 1-10 GeV (see \cite{Cohen:2009fz,Kribs:2009fy,Nussinov:1985xr,Barr:1991qn,Barr:1990ca,Gudnason:2006ug} for exceptions). 
 
Models of ADM typically assume that the universe has a net $B-L$ asymmetry \cite{Kaplan:2009ag}, generated by an unspecified baryogenesis mechanism at some high temperature.  This asymmetry is subsequently transferred to a dark sector, and this asymmetry in the dark  sector fixes the DM abundance.  For the asymmetry to dominate the energy density of the DM, the symmetric part must annihilate away efficiently.  It is not always straightforward to achieve a sufficiently high annihilation cross section.  After all, the DM is not charged under $U(1)_{EM}$ or $SU(3)_{C}$.  Furthermore, its light mass, when combined with constraints on the invisible width of the $Z^0$ boson, precludes large interactions via the weak force.    If a higher dimension operator is responsible for this annihilation, the suppression scale needs to be near or below the weak scale to achieve a large enough annihilation cross section \cite{Beltran:2010ww}.   Then one must ask the question why no hint of this new physics has been observed yet. Hence, the requirement of large symmetric annihilation cross sections implies a challenge for ADM model building.   

One possible solution to this problem occurs when light fields couple strongly to the DM.   For example, an axion from the next-to-minimal supersymmetric standard model (NMSSM) can play this role.  The DM can efficiently annihilate to the singlet axion which subsequently decays; this mechanism was employed in \cite{Kaplan:2009ag}.   Here, we build on this approach.  Suppose the dark sector contains a new dark force, and the dark gauge boson has a mass lighter than the DM, \emph{i.e.} roughly a GeV. Then the light gauge boson can provide the light annihilation mode, in analogy with the NMSSM axion. The cross section for this annihilation can be large, solving the challenge of reducing the symmetric component of the DM.  If the dark gauge boson has a small kinetic mixing with $U(1)_{Y}$, it can subsequently decay to SM fermions.  In addition, supersymmetrizing these models can provide ways for the sub-weak scale to be generated naturally \cite{Hooper:2008im,Feng:2008ya,ArkaniHamed:2008qp,Zurek:2008qg,Cheung:2009qd,Katz:2009qq,Morrissey:2009ur}. 

Large direct detection cross sections can result from the vector interaction in models where the DM annihilates to a $U(1)_{d}$ gauge boson that mixes with the SM photon.  DAMA \cite{Bernabei:2010mq} and the recent CoGeNT \cite{Aalseth:2010vx} results hint at a light DM candidate with a large cross section.  The mass of the DM required to explain these signals is in the correct range for ADM \cite{Fitzpatrick:2010em,Chang:2010yk,Kopp:2009qt,An:2010kc}.  
Whether or not these hints are borne out in future experiments, the models presented here demonstrate that the observation of light DM at direct detection experiments might point towards a model of GeV hidden sector ADM.

In the next section we present a toy model that illustrates the main features of ADM models with dark photons.  In Sec.~\ref{sec:SUSYmodel} we give a realistic supersymmetric (SUSY) model which realizes this paradigm.  In Sec.~\ref{sec:Cosmo} we discuss the cosmological history of this simple SUSY model.  In Sec.~\ref{sec:DD} we discuss the direct detection cross section and then turn in Sec.~\ref{sec:Collider} to exploring the collider implications of this model.  Then we conclude.  

\section{Ingredients}

In models of ADM, there are two key ingredients: an operator that transfers the asymmetry from the SM to the DM and a large annihilation mode that effectively suppresses the symmetric component of the relic density.  In this section we present a simple non-SUSY model that demonstrates the broad features of ADM models with a dark Abelian gauge group. 

The Lagrangian for the dark sector is
\bea
\mathcal{L} &=& \bar{\chi} (i \slashed{D} - m_\chi) \chi + |D_\mu H'|^2- V(H') \nonumber\\
&&- \frac{1}{4} b_{\mu\nu} b^{\mu\nu} + \frac{\epsilon}{2} b_{\mu\nu} B^{\mu\nu} + \op_\mathrm{asym}.
\eea
Here $b_{\mu \nu}$ and  $B_{\mu \nu}$ are the dark gauge boson and hypercharge field strengths, respectively.  $\chi$ is a Dirac fermion with charge $Q$ under $U(1)_d$ -- it is the DM, and $H'$ is the dark Higgs with charge $-1$ under $U(1)_d$. The operator $\op_\mathrm{asym}$ transfers the $B-L$ asymmetry from the SM sector to the dark sector.  $\epsilon$ parametrizes a kinetic mixing between the dark photon and the hypercharge boson.  It is naturally generated by integrating out matter charged under both symmetries; the result is an $\epsilon$ of the size \cite{Holdom:1985ag}:
\beq
\epsilon \sim \frac{g_Y g_d}{16\pi^{2}} \log{\frac{M'}{M}},
\eeq
where $g_Y$ is the hypercharge coupling constant; $g_d$ is the $U(1)_d$ coupling constant, and the logarithm of scales results from splittings between fields charged under both symmetries.  Due to the loop factor suppression, $\epsilon\sim 10^{-3}$, at least in the absence of large logarithmic enhancements.  When $H'$ acquires a non-zero vacuum expectation value (vev), the dark $U(1)$ is broken and the dark photon becomes massive.  The dominant symmetric annihilation mode for the DM is $\bar{\chi} \chi \to \gamma_d \gamma_d$.  

The asymmetry transfer operator must conserve dark charge, and so is of the form
\beq
\op_\mathrm{asym} = \frac{\left(H'^n \chi^p\right) \op_{B-L}}{\Lambda^r}
\eeq
where $\Lambda$ is the mass suppression scale, $p = n/Q$, and $\mathcal{O}_{B-L}$ is an operator with a non-zero $B-L$ number that involves only SM fields.  $p>1$ is a necessary condition for ensuring the stability of the DM.  Using the equilibrium methods outlined in \cite{Harvey:1990qw}, one can solve for the DM asymmetry in terms of the $B-L$ asymmetry.  If this asymmetric component dominates, the measured value of the DM relic density determines the mass of the DM.  
We discuss how the choice of transfer operator and corresponding $\Lambda$ singles out a DM mass in Sec.~\ref{sec:Cosmo}. Here, we note only that this operator need be in equilibrium after the baryon asymmetry is generated, but must go out of equilibrium before $T \sim m_{\chi}$, or the DM asymmetry will be Boltzmann suppressed.

\section{A Supersymmetric Model \label{sec:SUSYmodel}}

Supersymmetry will stabilize both the electroweak scale as well as the dark scale.  While in the above model the DM mass is put in by hand, here we can generate it dynamically.  We propose the following model:
\beq
\mathcal{L}_{d} \supset \int d^2 \theta \left(\lambda S T H' + \frac{\epsilon}{2} \mathcal{W}_d \mathcal{W}_Y \right).  \label{eq:superpotential}
\eeq
Here $S$ is a singlet, while $T$ has charge $+1$ under $U(1)_d$.  The dark Higgs, $H'$, has charge $-1$ under $U(1)_d$.  $\mathcal{W}_d$ and $\mathcal{W}_Y$ represent the gauge field strength superfields for the dark photon and hypercharge, respectively, with kinetic mixing  $\epsilon$.  In the absence of large soft terms in the hidden sector, this model gives rise to a symmetry breaking pattern where $\vev{S}=\vev{T}=0$ and $\vev{H'}\neq 0$ \cite{Cheung:2009qd,Morrissey:2009ur}.\footnote{Note this superpotential was also recently considered in an attempt to explain the CoGeNT excess in \cite{Essig:2010ye}, in a symmetric DM model and with different assumptions about supersymmetry breaking.}   
There is an accidental global symmetry under which $S=+1$ and $T=-1$, leading to a stable state.  The lightest component of the $S$ and $T$ chiral superfields constitutes the DM.  

We suppose SUSY breaking is communicated to the MSSM by gauge mediation, while the $U(1)_d$ does not couple directly to the messengers.  Then the hidden sector is shielded from SUSY breaking in the MSSM and only receives soft-terms via the small kinetic mixing parameter.   Once electroweak symmetry is broken, the kinetic mixing induces an effective Fayet-Illiopoulos (FI) $D$-term  for the $U(1)_{d}$, $\epsilon \langle D_Y \rangle$,
as in \cite{Cheung:2009qd}.   Ignoring the small supersymmetry breaking effects, the potential is 
\beq
V =\frac{1}{2} \left( g_d( |T|^2 - |H'|^2) + \epsilon \langle D_Y \rangle \right)^2 + |\lambda|^2 \left(|S|^2|H'|^2+|S|^2|T|^2+|T|^2|H'|^2\right),
\eeq
where $\langle D_Y \rangle = \frac{g_Y v^2 c_{2\beta}}{4}+\xi_Y$.  
Here, $v=246\mbox{ GeV}$ is the effective MSSM Higgs vev, $\tan{\beta} = v_u/v_d$.  $\xi_Y$ is a ``fundamental" FI term for hypercharge whose existence is more model dependent.   For example, a weak scale $\xi_Y$ can be naturally generated in  $U(1)$ messenger models of gauge mediation \cite{Dimopoulos:1996yq}.   For $c_{2\beta} = -1$ and $\xi_Y = 0$, $\sqrt{|D_Y|} \simeq 72 \mbox{ GeV}$. 
Then for $\epsilon = 10^{-3}$ and $\epsilon \langle D_Y \rangle \simeq 5 \mbox{ GeV}^2$, the GeV scale has been generated from the weak scale. 
The dark Higgs, $H'$, obtains a vev to cancel the $D$-term
\beq
\vev{H'} = \sqrt{\frac{\epsilon\langle D_Y \rangle}{g_d}},
\eeq
from which the scalars obtain masses
\beq
m_{H'}^2= 2 g_d^2 \vev{H'}^2;~~~~~~~m^2_S = m^2_{T} = \lambda^2 \vev{H'}^2.
\eeq
The mass of the dark photon is 
\beq\label{eq:mGammaD}
m_{\tilde{\gamma}_d} = \sqrt{2}g_d \vev{H'}.
\eeq
At this point, the vacuum is supersymmetric.  The mass matrix in the fermion sector (in the $(\tilde{\lambda}_d,\tilde{H}',\tilde{S},\tilde{T})$ basis) is given by
\beq
{\cal M} = 
\left(
\begin{array}{cccc}
0 & \sqrt{2} g_d \vev{H'} &0 & 0 \\
\sqrt{2} g_d \vev{H'} & 0 & 0 &0 \\
0& 0 & 0 & \lambda \vev{H'}  \\
0 & 0 & \lambda \vev{H'} & 0
\end{array}
\right).
\eeq
The dark Higgsino-photino mass eigenstate, $\tilde{\gamma}_d$, is degenerate with the gauge boson, and the $S-T$ fermions, $\psi$ and $\bar{\psi}$, are degenerate with their scalar superpartners.  

We now address how the small SUSY breaking effects leak into this sector.  In particular, two loop gauge mediated diagrams contribute positive mass squareds to the $T$ and $H'$ scalars via the kinetic mixing.  We normalize the size of this contribution to right handed selectron mass, $\tilde{m}_{E^c}$, as
\beq\label{eq:TMassCorrec}
\Delta \tilde{m}_{T,H'}^2 = \epsilon^2 \left(\frac{g_d}{g_Y}\right)^2 \tilde{m}_{E^c}^2.
\eeq
This equation is valid at the messenger scale; renormalization group running to the hidden sector scale is a 10\% effect.  This soft mass affects the cosmology of this model since it raises the $T$ scalar above $\psi$.  

Because it is a singlet, the $S$ scalar does not receive a positive (mass)$^{2}$ from gauge mediation.  Rather, it has a negative soft mass squared at one loop due to the presence of the $T$ and $H'$ soft masses.  This lowers the $S$ scalar just below $\psi$ by an amount
\beq\label{eq:SMassCorrec}
\Delta \tilde{m}_{S}^2 = - \frac{2 \lambda^2}{16 \pi^2} (\Delta\tilde{m}_{H'}^2+\Delta\tilde{m}_T^2) \log\left(\frac{M_\mathrm{mess}}{m_{S}}\right).
\eeq
Here $M_\mathrm{mess}$ is the messenger scale where the soft masses are generated. Thus the lightest state charged under the $S/T$ parity is the $S$ scalar. It is this state which constitutes the DM.

While the splittings of Eqs.~(\ref{eq:TMassCorrec}) and (\ref{eq:SMassCorrec}) will be most important for cosmology, for completeness we note the leading splitting in the gauge multiplet.  The dark photino gets a small correction from mixing with the MSSM gauge sector that splits the fermion into two Majorana states around the dark gauge boson.  Including the leading corrections to the dark photino mass,
\beq
m_{\tilde{\gamma}_d}^{(1,2)} = \sqrt{2}g_d \vev{H'} \pm \epsilon^2 \left(\frac{m_Z^2 s_W^2 s_{2\beta}}{\mu} + \frac{m_{\tilde{\gamma}_d}^2}{M_1} \right).
\eeq
There are two contributions to the mass of the dark Higgs radial mode, $h'$, which take it away from the SUSY limit: the small correction  from mixing with the Higgs boson via the $D$-term and a 1-loop radiative correction which contributes to its quartic.  The correction to the quartic is the larger of the two.  It shifts the physical dark Higgs boson mass by an amount 
\beq
\Delta{m}_{h'}^2 = \frac{\lambda^4 \vev{H'}^2}{16\pi^2} \log{\frac{m_T^2}{m_\psi^2}} \simeq  \frac{\lambda^2}{8\pi^2}   \Delta \tilde{m}_T^2.
\eeq

\begin{figure}
  \begin{center}
    \setlength{\unitlength}{10 cm}
 \begin{picture}(0.7,0.4)
     \thicklines
     \put(0.3,0.44){\line(1,0){0.3}} %T state
     \put(0.2,0.4){\line(1,0){0.4}} %psi state
     \put(0.3,0.38){\line(1,0){0.3}} %S state
     \put(-0.035,0.39){$\sim 10 \mbox{ GeV}$} %
     \put(0.7,0.44){$T$}
     \put(0.7,0.4){$\psi$}
     \put(0.7,0.35){$S$}
     \put(0.2,0.25){\line(1,0){0.4}} %vector boson state
     \put(0,0.24){$\sim \mbox{ GeV}$} %gauge state
     \put(0.7,0.25){$\gamma_d,\tilde{\gamma}_d,H'$}
     \put(0.2,0.1){\line(1,0){0.4}} %Gravitino state
     \put(0,0.09){$\ll \, \mbox{GeV}$} %
     \put(0.7,0.09){$\tilde{G}$}
   \end{picture}
  \end{center}
  \caption{The spectrum of the SUSY model.  We have illustrated the mass pattern of the $S/T$ multiplet (not to scale) since this splitting determines the identity of the DM.  The splittings within the dark photon multiplet have been suppressed.} \label{fig:SXYspectrum}
\end{figure}
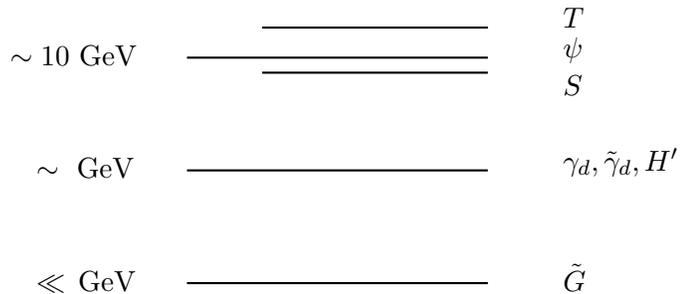

To allow efficient annihilation of the $S/T$ sector to gauge bosons, we choose $\sqrt{2} g_{d} < \lambda$. The spectrum is shown schematically in Fig.~\ref{fig:SXYspectrum}.  Aside from the gravitino, $\tilde{\gamma}_d$ is the lightest R-odd particle.    Although the dark gaugino is slightly lighter than the gauge bosons, thermal effects allow it to annihilate to the gauge bosons which subsequently decay.  We describe this process in detail in Sec.~\ref{sec:Cosmo}. 

There are phenomenological constraints on an Abelian GeV hidden sector.  If the dark photon mass is smaller than the mass of the $\Upsilon(3s)$, the lack of observation of dark photons at $B$-factories constrains the $m_{\gamma_d}-\epsilon$ parameter space \cite{Essig:2009nc}, yielding $\epsilon \lesssim 4\times 10^{-3}$.  For larger dark photon masses, the strongest constraints are $\epsilon \lesssim 10^{-2}$ coming from precision electroweak measurements -- there are $\epsilon$ suppressed couplings to the $Z^0$ which can lead to changes in these observables \cite{Gopalakrishna:2008dv}.  Finally, avoiding Landau poles for $\lambda$ before the GUT scale enforces $\lambda \lesssim 1.5$ which (due to the requirement that $\sqrt{2} g_d < \lambda$ in our model) constrains $g_d \lesssim 1.1$.  If one only requires no Landau poles appear before $\mathcal{O}(10\mbox{ TeV})$, this constraint is $\lambda \lesssim 2.5$ and $g_d \lesssim 1.8$.  Stronger constraints on $\epsilon/g_d$ from the Landau pole are dependent upon the DM mass (see Sec.~\ref{sec:Cosmo} below).  We plot the excluded region due to all of these constraints in Fig.~\ref{fig:photinosBBN}.

%%%%%%%%%%%%%%%%%%%%%%%%%%%%%%%%%%%%%%%%%%%%%%%%%%%%%%%%%%%%%%%%%%%%%%%%%%%%%%%%%
\section{Cosmology \label{sec:Cosmo}}
%%%%%%%%%%%%%%%%%%%%%%%%%%%%%%%%%%%%%%%%%%%%%%%%%%%%%%%%%%%%%%%%%%%%%%%%%%%%%%%%
The proposed SUSY model of ADM with a dark photon has a non-trivial cosmological history.  In particular, the near degeneracy of the states which comprise the DM and massive dark photon superfields imply the potential for late decays.  In the analysis that follows, we demonstrate that we maintain the success of Big Bang Nucleosynthesis (BBN).   In addition, the presence of the global symmetry on $S/T$ and R-parity results in two stable states, $S$ and the gravitino. We must check that $\Omega_\mathrm{DM}$ is dominated by the asymmetric part of the $S$ density.   

We present two different asymmetry transfer operators.  One of these operators has processes that re-symmetrize the DM and anti-DM at late times.
The model with a symmetric DM density today is subject to additional constraints.  For this reason, this transfer operator must be discussed separately.  

In all cases, we assume that the gravitino mass is $\leq 16 \mbox{ eV}$,  consistent with low-energy gauge mediation, in order to evade constraints from measurements of the Lyman-$\alpha$ forest without restricting the reheat temperature after inflation \cite{Viel:2005qj}.  We will conclude this section with some variations on our canonical cosmology.

One key component of ADM models is the requirement of an asymmetry transfer mechanism.  We assume that the transfer occurs via some higher dimensional operator, $\mathcal{O}_\mathrm{asym}$, generated by integrating out physics at a scale, $M$.  The states integrated out to generate $\mathcal{O}_\mathrm{asym}$ can be charged under both $U(1)_d$ and $U(1)_Y$, and in principle could also be responsible for generating $\epsilon$.  We are agnostic about the source of the $(B-L)$ asymmetry - we only require that it is generated before $\mathcal{O}_\mathrm{asym}$ falls out of equilibrium.

Since the $S$ field is a gauge singlet,\footnote{One is also free to use the combination $(TH')$ in constructing these operators.  } asymmetry transfer operators will have the following generic form \cite{Kaplan:2009ag}:  
\beq
\mathcal{O}_\mathrm{asym} = \frac{S^p \mathcal{O}_{B-L}}{M^r},
\eeq
where $\mathcal{O}_{B-L}$ is a SM gauge singlet operator involving only MSSM fields with a non-zero $B-L$ number $q$.  This operator transfers the $B-L$ into the $S/T$ global symmetry. The four lowest dimension MSSM superpotential operators with $|q|=1$ are $LH_u$, $U^c D^c D^c$, $LLE^c$, or $LQD^c$.  Higher $q$ operators can be built from combinations of these. The size of the asymmetry produced only depends on $q$ \cite{Chang:2009sv,Feldstein:2010xe}.

Assuming the symmetric component of the DM abundance is negligible (we verify this in specific cases below), we can compute the $S-S^\dagger$ asymmetry using standard equilibrium methods \cite{Harvey:1990qw}.  Above the $\vev{H'} \neq 0$ phase transition there is the additional requirement that the net $U(1)_d$ charge is zero. If $\mathcal{O}_\mathrm{asym}$ decouples before the electroweak phase transition (EWPT), the mass for the DM in the SUSY model is given by 
\beq
m_\mathrm{DM} = \frac{158}{33} \frac{p}{|q|} \frac{\Omega_\mathrm{DM}}{\Omega_B}\frac{B}{B-L} m_p \simeq (7.1 \textrm{ GeV}) \frac{p}{|q|}, \label{eq:mass}
\eeq 
where $m_p$ is the proton mass. $\Omega_\mathrm{DM}$ is the DM relic abundance, and $\Omega_B$ is the abundance of baryonic matter. $B/(B-L) \simeq 0.35$ with an uncertainty of $\mathcal{O}(10\%)$ due to the details of the sphalerons and the EWPT \cite{Harvey:1990qw}. If the asymmetry transfer operator decouples after the EWPT but before the dark sector phase transition (which occurs at $T \sim m_\mathrm{DM}$), the effective $B-L$ transferred is different, and 
\beq
m_\mathrm{DM} = \frac{197}{87} \frac{p}{|q|} \frac{\Omega_\mathrm{DM}}{\Omega_B}\frac{B}{B-L} m_p \simeq (3.3 \textrm{ GeV}) \frac{p}{|q|}. \label{eq:mass2}
\eeq 
In the main body of the text, we will focus on the operators:
\bea
\mathcal{O}_\mathrm{asym}^{(1)} &=& \frac{S^2 U^c D^c D^c}{M_{(1)}^2} \left(\mbox{ or } \frac{S^2 L L E^c}{M_{(-1)}^2}\mbox{, etc.}\right);\\
\mathcal{O}_\mathrm{asym}^{(-2)} &=& \frac{S^2 (L H_u)^2}{M_{(-2)}^3},
\eea  
where the superscript refers to the $B-L$ number, $q$, for the MSSM operator.  We will show in an appendix that $\mathcal{O}_\mathrm{asym}^{(-1)} = (S^2 L H_u)/M_{(-1)}$ does not give rise to a viable cosmology when all constraints are analyzed.  If they decouple above the EWPT, these operators imply a DM mass:
\bea
m_\mathrm{DM}^{(1)} = 14.2 \mbox{ GeV} &\Rightarrow& \lambda \sqrt{\frac{\epsilon/g_d}{10^{-1}}} \left(\frac{\sqrt{\langle D_Y \rangle}}{72\mbox{ GeV}}\right) =   0.62\label{eq:mDMqEq1};\\
m_\mathrm{DM}^{(-2)} = 7.1 \mbox{ GeV} &\Rightarrow& \lambda \sqrt{\frac{\epsilon/g_d}{10^{-2}}} \left(\frac{\sqrt{\langle D_Y \rangle}}{72\mbox{ GeV}}\right) =  1.0.
\eea
Hence, the choice of operator implies a relationship among the parameters in the dark sector.  One can use the upper bounds on $\lambda$ arising from the absence of a Landau pole to constrain the minimum allowed $\epsilon/g_d$, see Fig.~\ref{fig:photinosBBN}.

\subsection{After Decoupling of Asymmetry Transfer\label{sec:CosmoafterDecoupling}}

After $U(1)_d$ is broken, the asymmetric DM abundance is spread across $S,T,\mbox{ and } \psi$ in the ratios $\frac{1}{3}$, $\frac{1}{3}$, and $\frac{1}{3}$. However, the $\psi$ and $T$ are unstable.   Since we are working in the context of low scale gauge mediation, the decays $T \to \tilde{G} \psi$ and $\psi \to \tilde{G} S$ are allowed.  Decays to gauginos are kinematically forbidden due to the small mass splitting between the $S,~T$ scalars and $\psi$ fermion.  The decay width for these processes are:
\bea 
\label{eqn:decaytogravitino}
\Gamma(T \to \psi \tilde{G}) &=& \frac{1}{8\pi} \frac{(m_T^2 - m_\psi^2)^4}{F^2 m_T^3}; \\
\Gamma(\psi \to S \tilde{G}) &=& \frac{1}{16\pi} \frac{(m_\psi^2 - m_S^2)^4}{F^2 m_\psi^3}, 
\eea
where we have assumed a massless gravitino.  Since the decays are invisible to the SM, these processes will not interfere with BBN predictions.  In any case, for the parameters we consider, they occur on time scales less than a second.  These mass splittings are calculable in terms of the underlying parameters and are given by (see Eqs.~(\ref{eq:TMassCorrec}) and (\ref{eq:SMassCorrec})):
\bea
m_T^2 - m_\psi^2 &\simeq& 3\times 10^{-3} \left(\frac{g_d \epsilon}{10^{-4}}\right)^2\left(\frac{\tilde{m}_{E^c}}{200\mbox{ GeV}}\right)^2\mbox{ GeV}^2;\\
m_\psi^2 - m_S^2 &\simeq& 6\times 10^{-4} \lambda^2 \left(\frac{g_d \epsilon}{10^{-4}}\right)^2\left(\frac{\tilde{m}_{E^c}}{200\mbox{ GeV}}\right)^2\mbox{ GeV}^2.
\eea

Depending upon the asymmetry transfer operator, decays that change the DM asymmetry number by two units could also be allowed.  This `re-symmetrization' of the DM must occur when the DM number density is sufficiently low to prevent annihilations from turning back on, re-coupling the DM and reducing the relic density.  Since the cross section for annihilation of DM is large in these models, the operators that allow re-symmetrization of the DM abundance are also tightly constrained by indirect signals.  We will discuss this further when we consider specific asymmetry transferring operators.

The symmetric abundance of $S$ should be subdominant to the asymmetric density, so that the DM density is truly set by the baryon asymmetry and not thermal freeze-out.  The $S$ annihilations are dominated by the process $SS^\dagger \rightarrow \tilde{\gamma}_d \tilde{\gamma}_d^\dagger$, which comes from the $t$-channel exchange of a $T$ fermion. This annihilation cross section is approximately
\beq
\vev{\sigma_\mathrm{sym} v} \simeq \left(2 \times 10^{-20} \textrm{cm}^3/\textrm{s} \right)\lambda^4 \left(\frac{7 \textrm{ GeV}}{m_S}\right)^2,
\label{Sann}
\eeq
where we have assumed that the gauge sector is much lighter than the ADM sector.  This yields a {\em symmetric} relic density of 
\beq
\Omega_S^\mathrm{sym} h^2 \simeq 2\times10^{-8} \lambda^{-4}\left(\frac{m_S}{7 \textrm{ GeV}}\right)^2 \ll 0.1,
\eeq
which is clearly subdominant to the measured abundance of DM.

\begin{figure}[t]
\centering
\includegraphics[width=0.45\textwidth]{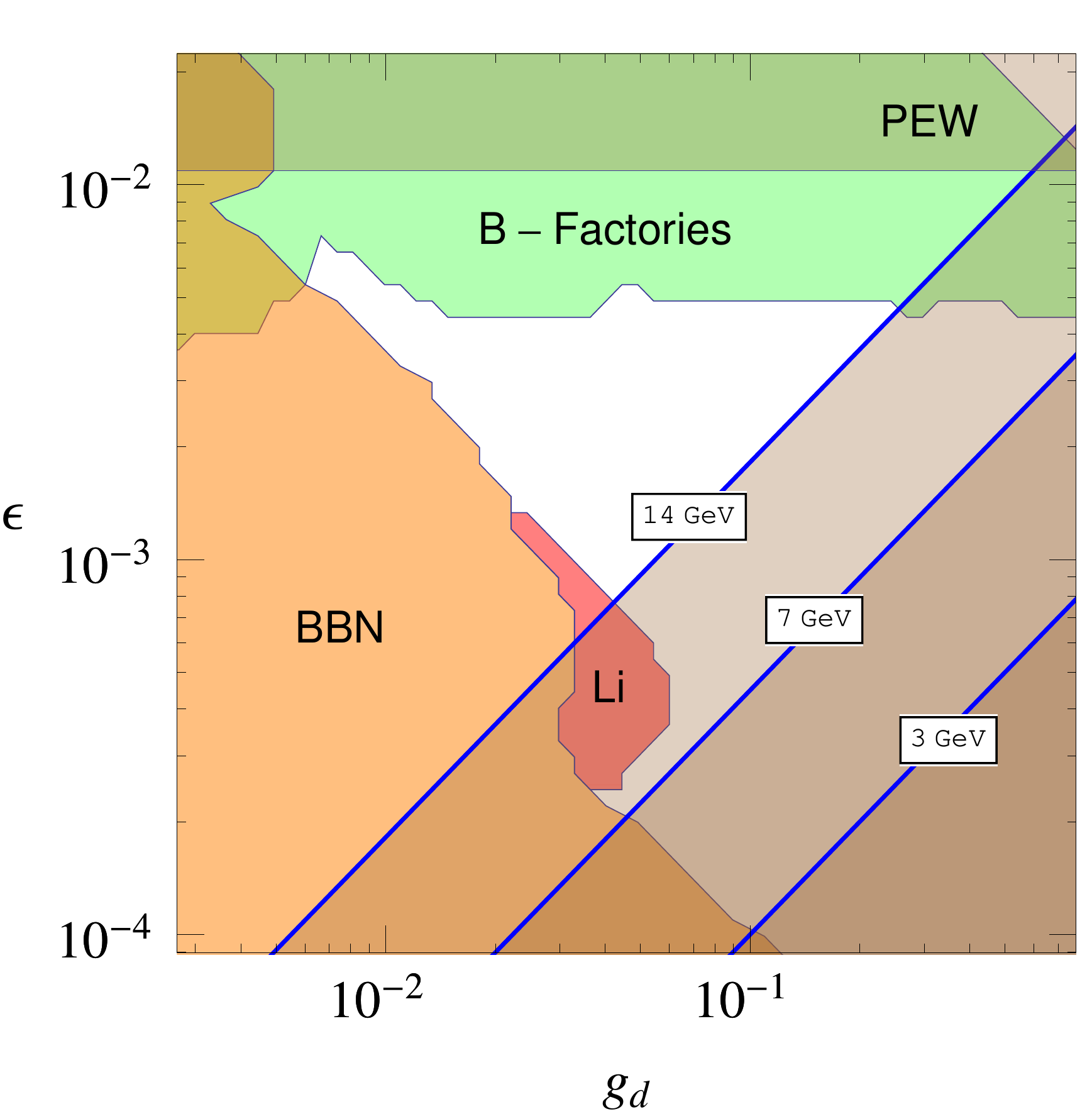}
\includegraphics[width=0.463\textwidth]{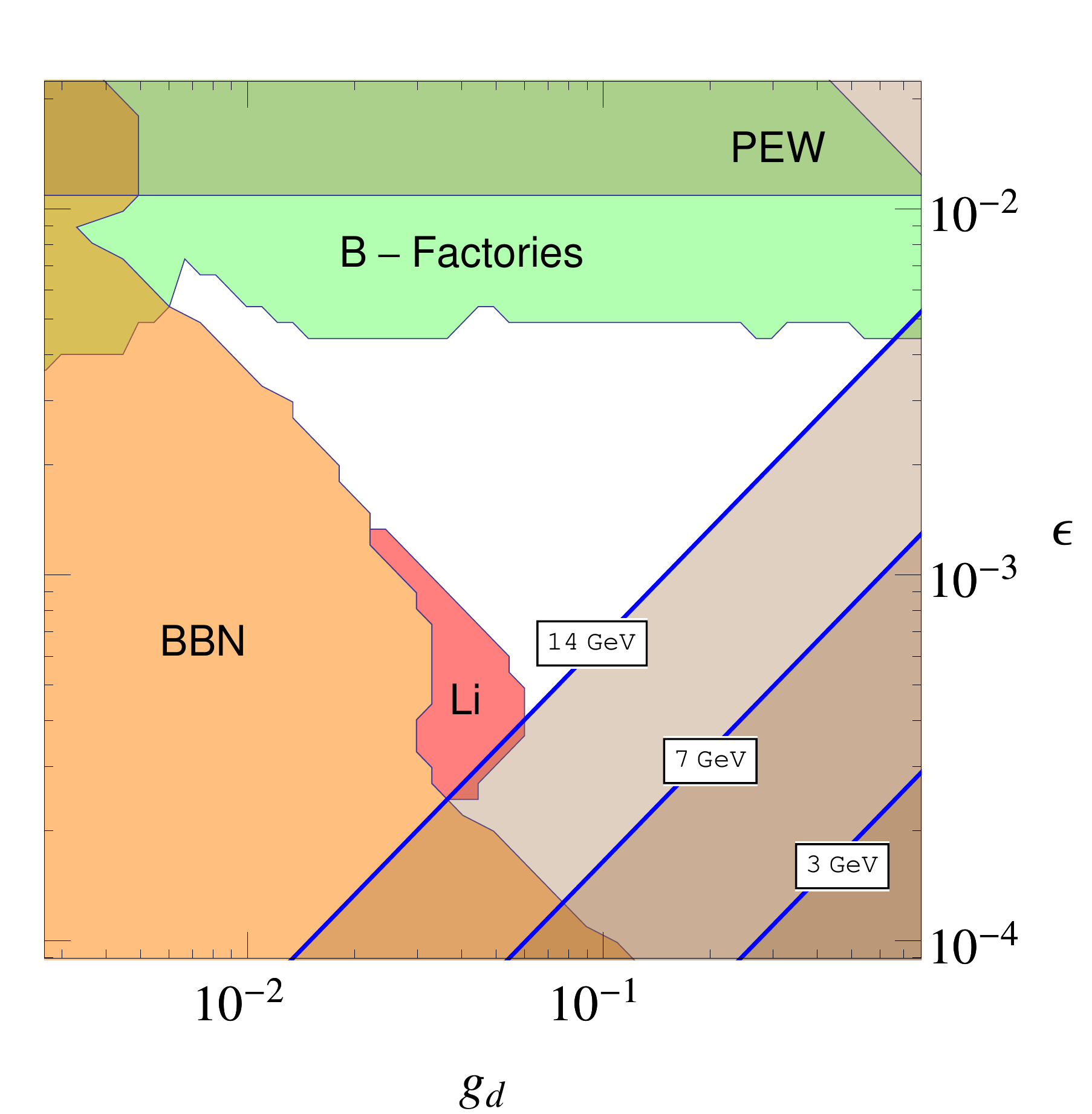}
\caption{Constraints in the $\epsilon-g_d$ plane.  We have shown the regions which are excluded by BBN constraints due to $\tilde{\gamma}_d\rightarrow \gamma \tilde{G}$ \cite{Jedamzik:2006xz} (orange), $B$-factories due to direct searches for $\gamma_d$ \cite{Essig:2010ye} (green), and precision electroweak measurements due to $\gamma_d-Z^0$ mixing \cite{Gopalakrishna:2008dv} (brown).  The red region corresponds to parameters which solve the lithium-7 problem \cite{Jedamzik:2006xz}.  One the left (right) we show contours where $\lambda$ is constrained so as not to reach Landau pole before $M_\mathrm{GUT}$ (10 TeV) for $m_\mathrm{DM} = 14.2\mbox{ GeV}$,  $m_\mathrm{DM} = 7.1\mbox{ GeV}$ and  $m_\mathrm{DM} = 3.3\mbox{ GeV}$, assuming $\vev{D_Y} = 72$ GeV.  The region below these contours is excluded.}
\label{fig:photinosBBN}
\end{figure}

The cosmology of $\gamma_d$ and $h'$ is straightforward since they both decay to the SM via $\epsilon$ suppressed couplings long before BBN.  The story is not so simple for the dark photino.  The presence of R-parity stabilizes the lightest of the superpartners, which for this scenario (low energy SUSY breaking), is the gravitino.  The dark photino is the second lightest R-odd state, and decays via $1/F$ suppressed couplings.  Due to the dark photino's near degeneracy with the dark photon, the dominant decay channel is $\tilde{\gamma}_d \to \gamma \tilde{G}$, which is suppressed both by the scale SUSY breaking and the kinetic mixing $\epsilon$.  This decay time is \cite{Cheung:2009qd}
\beq
\tau(\tilde{\gamma}_d \to \gamma \tilde{G}) = 190 \textrm{ s} \left( \frac{10^{-3}}{\epsilon}\right)^2 \left( \frac{\mbox{GeV}}{m_{\tilde{\gamma}_d}}\right)^5 \left(\frac{\sqrt{F}}{50\mbox{ TeV}}\right)^4.
\eeq
This late production of photons could, in principle, alter the predictions of BBN.  This depends on the destructive power of the dark photinos, which is given by $m_{\tilde{\gamma}_d}
n_{\tilde{\gamma}_d}/s \equiv m_{\tilde{\gamma}_d} Y_{\tilde{\gamma}_d}$,
where $n_{\tilde{\gamma}_d}$ is the number density of photinos and $s$ is the entropy density of the universe.   Since the Higgsino component of the dark photino induces an interaction between the dark photino and the dark photon, the number density is set by these interactions.  Though the dark photino and photon masses are degenerate, the thermal tail of the Boltzmann distribution allows efficient annihilation of the dark photinos.  To good approximation, the annihilation cross-section for this process is given by \cite{Morrissey:2009ur}:
\beq
\langle \sigma_{\tilde{\gamma}_d} v \rangle \simeq \frac{g_d^4}{16 \pi m_{\tilde{\gamma}_d}^2} v_{f.o.}
\simeq 7 \times 10^{-24} \mbox{cm}^3/\mbox{s}\left(\frac{g_d}{0.1}\right)^4\left(\frac{1 \mbox{ GeV}}{m_{\tilde{\gamma}_d}}\right)^2 \left(\frac{v_{f.o.}}{0.3}\right), 
\label{gammadann}
\eeq
where $v_{f.o.}$ is the velocity when the dark photinos freeze out.  Hence, the dark photinos can have a small relic abundance when they decay to a gravitino and a photon. 
In Fig.~\ref{fig:photinosBBN} we show the regions in the $g_d-\epsilon$ plane which do not alter the predictions of BBN  and satisfy constraints from $B$-factories and from precision electroweak (PEW) measurements.  In generating this figure we have done the full calculation of the thermally averaged cross section to capture the effects of the degeneracy between the initial and final states.  We also show the region of specific choices of $\epsilon$ and $g_d$ which can modify the abundance of Li-7, alleviating the tension with the current measurements \cite{Jedamzik:2006xz}.

Next we explore the cosmology associated with transferring the asymmetry to the DM.  We pay particular attention to the requirement that the transfer operator not imply a Boltzmann suppression for the asymmetry by remaining in  equilibrium to very low scales, $T  < m_\mathrm{DM}$. This requirement constrains the asymmetry transfer scale, $M$.  The physics involved in the determination of this scale is sensitive to the choice of the transfer operator, so we discuss each operator in turn.

\subsection{Cosmology of Models with $\mathcal{O}_\mathrm{asym}\sim S^2 U^c D^c D^c$}

The cosmology associated with the $q=1$ operator is the most straightforward.  Comments similar to those below also apply to operators where $U^{c} D^{c} D^{c}$ is replaced by either $LLE^c$ or $LQD^c$.  Since there are three MSSM fields involved which do not obtain vevs, at tree level all asymmetry transfer interactions will involve at least one SM superpartner.  For these processes the transfer rate will be Boltzmann suppressed for temperatures below the superpartner scale, and will be be strongly suppressed when $T\sim m_\mathrm{DM}$.  So, for low temperatures (below the SUSY scale), the dominant process arises from a one-loop diagram where a gluino is exchanged.  This coverts two squarks to quarks and generates an effective dimension-7 operator ($S \psi_{S} \psi_{d^c} \psi_{d^c} \psi_{u^c} /M_{eff}^{3}$).    Taking a superpartner scale of 1 TeV,  the requirement that this effective operator be out of equilibrium before $T \sim m_\mathrm{DM}$ enforces the mild constraint $M_{(1)} > 2$ TeV.

If one imposes the stronger bound that the transfer operator decouples before the EWPT, a stronger bound on $M_{(1)}$ is present. Depending on the precise spectrum of the superpartner masses, either the tree-level or loop induced process can be the most important.  However, both give bounds of  $M_{(1)}\sim \mathcal{O}(100\mbox{ TeV})$.  If this stronger condition holds, then the DM mass is as given in Eq.~(\ref{eq:mDMqEq1}), otherwise Eq.~(\ref{eq:mass2}) applies.

\subsection{Cosmology of Models with $\mathcal{O}_\mathrm{asym}\sim S^2 (L H_u)^2$}\label{sec:CosmoS2LH2}

For the $q=-2$ operator, the story is different:  the process $SS \rightarrow \nu^\dagger \nu^\dagger$ has the potential to wash-out the asymmetry.  Requiring that this process be out of equilibrium at temperatures of order the DM mass yields:
\beq\label{eq:M2Decoupling}
M_{(-2)} \gtrsim 20 \, \mbox{TeV} \, \left(\frac{m_S}{7\mbox{ GeV}}\right)^{1/6}.
\eeq
The mass estimate of $m_\mathrm{DM}$ in Eq.~(\ref{eq:mass}) requires the stronger condition that the asymmetry transfer operator decouples at temperatures above the EWPT (and does not recouple once $\vev{H_u} \neq 0$).  This implies that $M_{(-2)} \gtrsim 30 \mbox{ TeV}$.   

\begin{figure}[t]
\begin{center}
\includegraphics[width=0.3\textwidth]{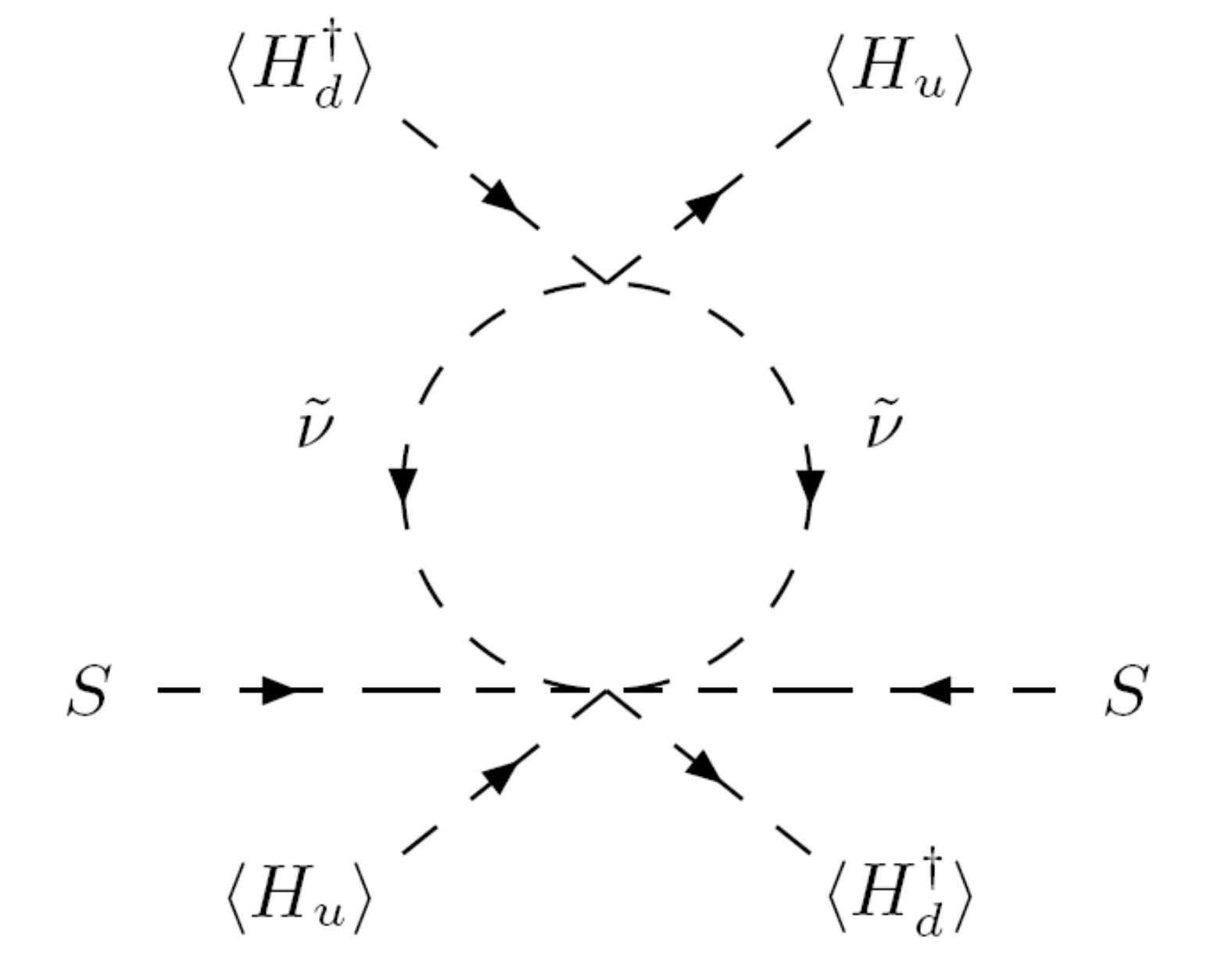}
\end{center}
\caption{The one-loop diagram which generates the $S$ number violating mass $b_S$.}
\label{fig:bSOneLoop}
\end{figure}

The origin of neutrino masses has a strong impact on the cosmology for this transfer operator. If neutrinos are Majorana, then the superpotential operator $(LH_u)^2/M_{\nu_R}$ is non-vanishing, where $M_{\nu_R}$ is the right handed neutrino mass scale. The operator $S^{2} (L H_{u})^2$ equates $L$ number with $S$ number.  Therefore, the neutrino mass operator violates $S$ number and generates a mass term via the one-loop diagram in Fig.~\ref{fig:bSOneLoop} that breaks $S$ number by two units, $b_S S S+\mathrm{h.c.}$.  This splits the real and imaginary components of the $S$ scalar by
\bea
\Delta m_S &=& \frac{b_S}{m_S} \simeq \frac{1}{16 \pi^2} \frac{v^2 c_{\beta}^2 \mu^2}{M_{(-2)}^3} \frac{m_{\nu}}{m_S}\log\left(\frac{\tilde{m}_{\nu_L}}{M_\mathrm{mess}}\right) \nonumber\\
&\simeq& 4\times 10^{-22} \mbox{ GeV} \; \left(\frac{7\mbox{ GeV}}{m_S}\right)\left(\frac{\mu}{100\mbox{ GeV}}\right)^2\left(\frac{10^5\mbox{ GeV}}{M_{(-2)}}\right)^3.
\eea

Here $\mu$ is the supersymmetric Higgs mass parameter; $\tilde{m}_{\nu_L}$ is the sneutrino soft mass, and $m_{\nu}$ is the neutrino mass.  This splitting will induce $S-S^\dagger$ oscillations when $H \sim \Delta m_S$ similar to \cite{Cohen:2009fz}.  

When these oscillations begin, one must check that the now symmetric relic density does not recouple and annihilate away.  This condition is given by $H(T_\mathrm{r}) > \Delta m_S$, where the re-coupling temperature $(T_\mathrm{r}\sim m_S^3/\lambda^4)$ is in the range $0.1-100 \mbox{ keV}$.  This constraint implies a limit on $M_{(-2)} \gtrsim 10^5 \mbox{ GeV}$ which is more restrictive than the decoupling constraints described above in Eq.~(\ref{eq:M2Decoupling}).  Hence, the asymmetry operator decouples before the EWPT, and the DM mass is $7.1 \mbox{ GeV}$, from Eq.~(\ref{eq:mass}). 

Even if the oscillations do not occur so early as to affect the relic density, they could lead to residual annihilation which could give additional constraints.  The annihilation mode $S S^\dagger \rightarrow \tilde{\gamma}_d^\dagger \tilde{\gamma}_d \rightarrow \gamma \gamma \tilde{G} \tilde{G}$ could produce photons which can effect the reionization depth of the CMB, see \cite{Slatyer:2009yq} for a recent analysis. The quantity constrained is the annihilation cross section times the ionization fraction, $f$.  For DM in the 10 GeV range, 
\bea\label{eq:CMBConstraint}
f \vev{\sigma v} \left(\frac{\rho_S}{\rho_\mathrm{DM}}\right)\left(\frac{\rho_{S^\dagger}}{\rho_\mathrm{DM}}\right) &=&  \frac{1}{4} f \vev{\sigma v}
\lesssim \mbox{few} \, \times 10^{-26} \mbox{ cm}^3/ \textrm{s}.
\eea
We expect that $f$ will be in the range $0.1-0.5$.  Hence, if oscillations occur before recombination, the requirement $\vev{\sigma v} \lesssim 10^{-24}$ translates to $\lambda \lesssim 0.1$. 

After fixing $\lambda \sim 0.1$, one must check that $m_S = 7.1\mbox{ GeV}$ can be achieved in this model.  To obtain a DM mass of this size, one must maximize the ratio $\epsilon/g_d$.  From Fig.~\ref{fig:photinosBBN}, the maximum this ratio can be is $(\epsilon/g_d)_\mathrm{max} \sim (7\times 10^{-3}/7 \times 10^{-3})$, which, when combined with $\lambda \sim 0.1,$ implies $m_S = 7.2 \mbox{ GeV}$.  
Hence, this scenario is marginally feasible.  Including a bare FI term for hypercharge could mitigate this tension.  Note that this point in parameter space should be probed by the existing but as yet unanalyzed data from the $B$-factories.

Alternatively, if $M_{(-2)} \gtrsim 10^{10} \mbox{ GeV}$, the oscillations occur at temperatures below an eV.  Hence there are no DM annihilations during recombination.  In this case, the strongest constraints come from considering the effect of DM annihilation on reionizing the universe.  Since the high energy photons which result from the $S S^\dagger \rightarrow \tilde{\gamma}_d^\dagger \tilde{\gamma}_d \rightarrow \gamma \gamma \tilde{G} \tilde{G}$ annihilations are poor ionizers \cite{Belikov:2009qx},
the strongest constraint comes from (for example) the annihilation channel $S S^\dagger \rightarrow \gamma_d \gamma_d \rightarrow e^+ e^- e^+ e^-$ where the electrons subsequently upscatter CMB photons.  This cross section is roughly two orders of magnitude smaller then the one quoted in Eq.~(\ref{Sann}).  This translates into a bound\footnote{In \cite{Belikov:2009qx} the DM mass is 100 GeV.  Since our DM is 7.1 GeV in this model the constraint will be slightly stronger then what they quote.} $\lambda \lesssim 0.3$ \cite{Belikov:2009qx}.  Note that this larger value of $\lambda$ will alleviate some of the tension with achieving the correct size for $m_S$.  

If $M_{-2}\gtrsim 10^{12}$ GeV, then DM has not begun oscillating yet.  Alternately, since the mass splitting is proportional to the Majorana neutrino mass, if the neutrinos have Dirac masses no oscillation occurs.  In these cases, the DM abundance would still be asymmetric today and the above constraints do not apply.

\subsection{Variations on the Cosmological History} 
In this section we will explore various other allowed cosmological histories beyond the simplest story we have presented above.  For example, one could imagine a scenario with a heavier gravitino.  The dark sector will generically feel anomaly mediated supersymmetry breaking contributions, which for too large a gravitino mass could potentially raise the DM above the GeV scale or destabilize the $H'$ vev.  This implies 
\beq
\frac{\alpha}{4\pi} m_{3/2} = \frac{\alpha}{4\pi} \frac{F}{\sqrt{3}m_\mathrm{Pl}} \lesssim \mbox{GeV} \Rightarrow \sqrt{F} \lesssim 2\times10^{10} \mbox{ GeV},
\eeq
which implies a bound of $m_{3/2} \lesssim 130 \mbox{ GeV}$.  Thus, the gravitino can be heavier than the dark photino.  In this case, the photino cannot decay, so one should ensure that the abundance of dark photinos is small enough to only constitute a subdominant portion of the DM:  
\beq
\Omega_{\tilde{\gamma}_d} h^2 \simeq 3.5\times 10^{-2} \left(\frac{0.02}{g_d}\right)^4 \left(\frac{m_{\tilde{\gamma}_d}}{1\mbox{ GeV}}\right)^2\left(\frac{0.3}{v_{f.o.}}\right).
\eeq
This implies a lower bound on $g_d \gtrsim 0.02$.  Alternately, a small amount of R-parity violation (RPV) in the MSSM could allow dark photino decays without spoiling BBN.  If this RPV is provided by a $LLE^c$ or $LQD^c$ operator, assuming no non-trivial textures, this implies a value for the coefficient near the limits from $\mu \to e \gamma$.

In this scenario, the gravitino would decay to dark photinos as well.  Again, constraints from BBN would limit the abundance of gravitinos produced in the early universe, which translates into a constraint on the reheat temperature of the universe of $\mathcal{O}(10^5\mbox{ GeV})$ \cite{Kawasaki:2008qe}.  This could pose a problem for asymmetry transfer operators which require higher reheat temperatures to ensure the transfer is ever in equilibrium.

Another way to avoid a gravitino overabundance is to imagine a too-large baryon asymmetry was generated via the Affleck-Dine mechanism \cite{Affleck:1984fy}, which was subsequently diluted by a period of late-time inflation to the measured value while simultaneously diluting the gravitinos.  Since the DM is set by the same large asymmetry, it would be diluted by the same fraction and would maintain the correct ratio between the relic density of baryons and DM. 

\section{Direct Detection}\label{sec:DD}
Since $S$ is neutral under the dark gauge force, tree-level direct detection proceeds either by the exchange of $h'$  via mixing with the MSSM Higgs, which is suppressed by $\epsilon$, or by mixing with the $T$ via the $A_\lambda$ term to exchange a dark photon.  However, since we have assumed that the only SUSY breaking is communicated to the dark sector through kinetic mixing, this $A$-term is suppressed by $\epsilon^2$.  So, these tree-level diagrams are small.  However, once $H'$ acquires a vev the $S$ scalar receives a coupling to the dark photon at the one-loop level:
\beq
\frac{\lambda^2 g_d}{16 \pi^2}\left(\frac{4 g_d^4-\lambda^4+4 \lambda^2 g_d^2 \mathrm{ log}\left(\frac{\lambda^2}{2 g_d^2}\right)}{2(2 g_d^2-\lambda^2)^{2}}\right) S^\dagger \overleftrightarrow{\partial_{\mu}} S \gamma_d^{\mu}\equiv g_d q_\mathrm{eff} S^\dagger \overleftrightarrow{\partial_{\mu}} S \gamma_d^{\mu}.
\eeq
This coupling is analogous to the one-loop $Z^0 b\bar{b}$ vertex corrections from a charged Higgs and a top quark \cite{Haber:1999zh}.

Since the dark photon only couples to the atomic number of the nucleus, the effective cross section per proton is 
\beq
\sigma_{p} = \frac{4}{\pi} \frac{g_W^4 c_W^4 \mu_{S,p}^2}{c_{2\beta}^2 m_W^4} q_\mathrm{eff}^2,
\eeq
where $g_W$ is the weak coupling constant, $c_W$ is the cosine of the weak mixing angle, and $\mu_{S,p}$ is the reduced mass of $S$ and a proton.  Due to the $g_d$ and $\epsilon$ dependence in $m_{\gamma_d}$ (see Eq.~(\ref{eq:mGammaD})), this cross section is approximately independent of both parameters \cite{Cheung:2009qd}.  This is
\beq
\sigma_{p} \simeq (9.1 \times 10^{-42} \mbox{cm}^2) \lambda^4 ,
\eeq
where we have taken the limit $\lambda \gg g_d$.  This is not large enough to give rise to the signal observed by CoGeNT, but it will be probed by a variety of upcoming experiments.

\begin{figure}[t]
\begin{center}
\includegraphics[width=0.7\textwidth]{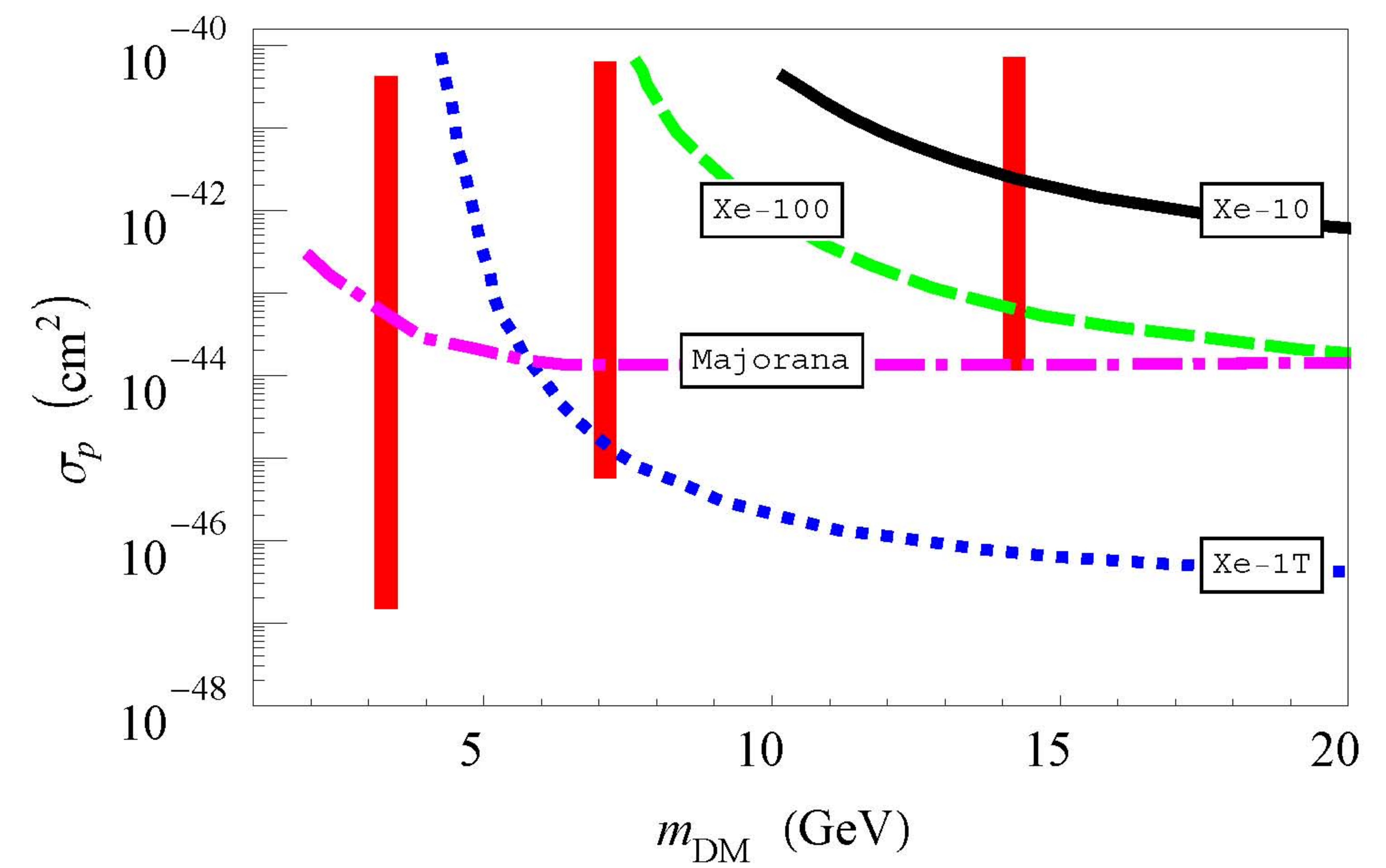}
\end{center}
\caption{The predictions for the direct detection scattering cross sections normalized per proton ($\sigma_p$) for $m_\mathrm{DM} = 14.2 \mbox{ GeV}$, 7.1 GeV and 3.3 GeV.  We have plotted current/projected limits (also normalized per proton) from Xenon-10 (solid black line), Xenon-100 with 6,000 kg-days (dashed green line),  Xenon-1T (dotted blue line) \cite{DMPlotter}, and Majorana (dot-dashed purple line) \cite{MAJORANA}.}
\label{fig:DirectDetection}
\end{figure}

In Fig.~\ref{fig:DirectDetection}, we have plotted the predicted range of direct detection cross sections, appropriate to $m_\mathrm{DM} = 14.2 \mbox{ GeV}$, 7.1 GeV and 3.3 GeV.  The upper bound is due to the assumption that there is no Landau pole for $\lambda$ before the GUT scale, and the lower bound occurs for the smallest allowed value of $\lambda$ consistent with the correct DM mass (using $(\epsilon/g_d)_\mathrm{max} \simeq 1$ as described in Sec.~\ref{sec:CosmoS2LH2}).  We also show the current Xenon-10 bound (solid black line), the projected Xenon-100 bound, assuming 6000 kg-days (dashed green line), the projected Xenon-1T bound (dotted blue line) \cite{DMPlotter}, and the projected limit from the Majorana experiment (dot-dashed purple line) \cite{MAJORANA}.  We have normalized these bounds by the factor $(Z/A)^2$ which is appropriate for our model where scattering is only off of protons ($\sigma_p$).  For $m_\mathrm{DM}$ =14.2 GeV, the largest values of $\lambda$ are already excluded by Xenon-10 \cite{Angle:2007uj}. At 14 GeV, the bound from Xenon-10 is approximately $3 \times 10^{-43}$ cm$^{2}$, which translates to $\lambda < 0.7$.  Nearly the entire parameter space for $m_\mathrm{DM} = 14.2 \mbox{ GeV}$ can be probed by Xenon-100 with 6,000 kg-days.  For $m_\mathrm{DM} = 7.1 \mbox{ GeV}$, Xenon-1T will cover the allowed region, and for $m_\mathrm{DM} = 3.3 \mbox{ GeV}$, Majorana will probe much of the allowed range.  Hence, a combination of current and proposed experiments will have the potential to cover most of the interesting parameter space for this model.  

\section{Colliders}\label{sec:Collider}
Finally, we  discuss some collider implications of this class of models.  There are three portals into the dark sector which could potentially be probed: photon kinetic mixing, Higgs boson mixing, and the asymmetry transfer operator.  

The MSSM LSP ($\mathrm{LSP}_\mathrm{MSSM}$) is unstable to decay to the low mass hidden sector \cite{Strassler:2006im,Strassler:2006qa}.  One mediation mechanism for decay to the hidden sector is through kinetic mixing, as discussed in \cite{Hooper:2008im,ArkaniHamed:2008qn}.  The collider phenomenology of such scenarios has been studied extensively recently; see for example \cite{Abazov:2009hn,Essig:2010xa,Schuster:2009au,Bjorken:2009mm,Essig:2009nc, Batell:2009yf,Batell:2009jf,Batell:2009di}.  

Photon kinetic mixing may also be probed via the decays of the $\mathrm{LSP}_\mathrm{MSSM}$ to the dark sector \cite{Hooper:2008im,ArkaniHamed:2008qp}.  If the $\mathrm{LSP}_\mathrm{MSSM}$ is has electroweak quantum numbers, then it will decay promptly to its SM partner and a dark gaugino via an $\epsilon$-suppressed interaction.  This dark gaugino is stable on detector time scales, and so will manifest as missing energy.  More interesting is if $\mathrm{LSP}_\mathrm{MSSM}$ is a neutralino, since it will decay to a dark gaugino and dark Higgs via $\epsilon$ mixing in the neutralino mass matrix.  The dark gaugino will again result in missing energy.  However, the dark Higgs will promptly decay back to SM fermions via mixing with the MSSM Higgs boson.  These could produce ``lepton jets'' \cite{ArkaniHamed:2008qp}.

The $T$ and $\psi$ fields couple to the $Z^0$ and the MSSM Higgs boson via $\epsilon$ suppressed couplings, so it will be difficult to produce these particles directly.  Furthermore, the DM state $S$ only interacts through couplings which are both $\epsilon$ and loop suppressed.  Hence, the LHC study of the DM will be indirect.  There will be rare decay of the Higgs boson either to a pair of dark photinos (invisible) or dark Higgs bosons (multijet). For the largest values of $\epsilon$ these branching ratios will be ${\mathcal O}$(10\%).

Finally, if the $\mathcal{O}_{B-L} \sim U^c D^c D^c$, then the UV completion will necessarily involve colored objects, some of which could have the quantum numbers of diquarks.  If this asymmetry operator decouples after the EWPT (which would imply a DM mass quoted in Eq.~(\ref{eq:mass2})), then this UV completion is a candidate for early discovery at the LHC \cite{Bauer:2009cc, Kang:2007ib}.

\section{Discussion and Conclusions}
In this work we have presented a supersymmetric model of Asymmetric Dark Matter, where the GeV scale for the DM mass is naturally generated by loop suppressed gauge kinetic mixing between the hypercharge and dark gauge bosons.  This scenario allows the symmetric component of the DM to annihilate efficiently into the dark photons.  Direct detection signals proceed via interactions with the dark photon.  

This model also provides a solution to the DM problem in models of low scale gauge mediation where the very light gravitino is the LSP and cannot account for the DM.  Since gauge mediation is a key component for achieving the appropriate spectrum in this model, the connection is robust.  Hence, we have shown that this ADM module can provide the DM for gauge mediated SUSY breaking models.

While we chose to focus on a simple model, this paradigm encompasses a large class of theories which connect the ADM mass scale with the weak scale via a one loop suppression.  We note two interesting mechanisms for achieving this goal.  The first was proposed in \cite{Zurek:2008qg,Morrissey:2009ur} and was coined ``singlet meditation" by these authors.  The idea is to mediate a weak scale soft mass to a hidden sector singlet field which is then transferred to the rest of the dark sector at one loop via Yukawa couplings.  Another choice uses the ideas of \cite{Katz:2009qq}, where the soft spectrum of the MSSM is due to gaugino (or gravity) mediation while the dark sector only receives contributions from anomaly mediation, again resulting in a one loop suppression.  Both of these ideas can be convolved with the ADM paradigm in straightforward -- if not minimal -- ways, resulting in an explanation for the GeV scale DM mass.  

Finally we note that the model presented here provides another example of GeV scale DM with an observable direct detection cross section.  The DM mass in models of this type are typically $\sim$ 10 GeV, so direct detection experiments with low energy thresholds are best suited to discover DM of this type.

\section*{Acknowledgements}
We thank A.~Adams, J.~Terning, C.~Csaki, M.~Papucci, J.~Thaler, T.~Volansky and I.~Yavin  for useful comments and discussions.  
This work was supported in part by NSF Career Grant NSF-PHY-0743315 (TC, AP) and by DOE Grant \#DE-FG02-95ER40899 (AP, DP).  
\def\theequation{\Alph{section}.\arabic{equation}}
\begin{appendix}

\setcounter{equation}{0}

\section{Models with $\mathcal{O}_\mathrm{asym}\sim S^2 L H_u$ are not allowed}\label{sec:CosmoS2LH}
In this appendix, we argue that this operator is excluded.  We begin by arguing for the allowed size of $m_\mathrm{DM}$ in models with this operator.

Since the size of $m_\mathrm{DM}$ is determined by when the asymmetry transfer decouples with respect to the EWPT, it depends on the size of $M_{(-1)}$.  In particular, the process $\psi \psi \leftrightarrow \nu^\dagger \tilde{\gamma}_d$, which proceeds via $t$-channel $S$ exchange, controls the transfer once $\vev{H_u}\ne0$.  Since the rate for this process is proportional to $T$ for $T_\mathrm{EWPT}>T>m_S$, it becomes more important as the temperature decreases.  Therefore, if this process were ever in equilibrium it would necessarily lead to some washout since its decoupling would be controlled by the Boltzmann suppression of the DM.  Requiring that this process not be in equilibrium for any $T>m_S$ implies a bound
\beq
M_{(-1)} \gtrsim 3 \times 10^{8} \mbox{ GeV} \left(\frac{\lambda}{0.1}\right) \left(\frac{14 \mbox{ GeV}}{m_S}\right)^{1/2}.
\eeq
For the asymmetry transfer to decouple before $T = T_\mathrm{EWPT}$, requires examination of the operator with $\vev{H_u}=0$, which gives the condition:
\beq
M_{(-1)}\gtrsim 6\times 10^7\mbox{ GeV} \left(\frac{\lambda}{0.1}\right).
\eeq
These two conditions together imply that in order to avoid washout, the asymmetry transfer must decouple at $T>T_\mathrm{EWPT}$, and the DM mass is 14.2 GeV (see Eq.~(\ref{eq:mass})).

For this operator, the decay $\psi \rightarrow S^{\dagger} \nu^\dagger$ is allowed.  This could give a non-trivial symmetric component of the DM today.   If this decay rate is sizable, the constraints described in Sec.~\ref{sec:CosmoS2LH2} are relevant which implies that $\lambda \lesssim 0.1$.  In fact, given the CMB constraint, it is not possible to achieve a DM mass as large as the required $14.2 \mbox{ GeV}$.  As described in Sec.~\ref{sec:CosmoS2LH2}, maximizing the ratio $\epsilon/g_d$ yields the largest possible DM mass.  Using Fig.~\ref{fig:photinosBBN}, this ratio attains its maximum at $(\epsilon/g_d)_\mathrm{max} \sim (7\times 10^{-3}/7 \times 10^{-3})$, which when combined with $\lambda \sim 0.1$ implies $m_S = 7.2 \mbox{ GeV}$.  Since this is far below 14.2 GeV, this scenario is excluded.   

One might hope that the CMB constraint could be mitigated by ensuring that symmetric decays $\psi \to \tilde{G} S$ dominate over the asymmetric decays.  The decay width to gravitinos is given in Eq.~(\ref{eqn:decaytogravitino}) and to neutrinos is given by
\beq\label{eq:PsiToSNuDecay}
\Gamma(\psi \rightarrow S^{\dagger} \nu^\dagger)=\frac{1}{32 \pi} \frac{ v^{2} \sin^{2} \beta}{M_{(-1)}^{2}}\frac{(m_\psi^2 - m_S^2)^{2}}{m_\psi^{3}}.
\eeq
Then the branching ratio is given by
\begin{equation}
\mathrm{BR}(\psi \rightarrow S \tilde{G}) = 1  - \mathrm{BR}(\psi \rightarrow S^{\dagger} \nu^\dagger)=\frac{ 2 M_{(-1)}^{2} (m_\psi^2 - m_S^2)^{2}}{F^{2} v^{2} s_\beta^2 + 2 M_{(-1)}^{2} (m_\psi^2 - m_S^2)^{2}}.
\end{equation}
To satisfy the CMB constraint for $\lambda =1$ requires that $\mathrm{BR}(\psi \rightarrow S^{\dagger} \nu) \lesssim 10^{-4}$.  For $g_d =10^{-1}$ and $\epsilon = 4\times 10^{-3}$, $M_{(-1)} \gtrsim 10^{16}\mbox{ GeV}$.  This implies the reheat temperature after inflation must be $\sim 10^{16}\mbox{ GeV}$ in order for this operator to ever have been in equilibrium, inconsistent with the lack of observation of tensor modes at WMAP \cite{Finelli:2009bs}.

\end{appendix}

\bibliography{ADMbiblio}  %For Bibtex

\end{document}